\documentclass[aps,prc,preprint,groupedaddress]{revtex4-1}

\usepackage{graphicx}

\begin{document}

\title{Meson spectra and $m_{T}$  scaling in p+p, d+Au and  Au+Au collisions at $\surd s_{NN} $ = 200 GeV }

\author{\large P. K. Khandai$^1$}
\author{\large P. Shukla$^{2,3}$}
\email{pshukla@barc.gov.in}
\author{\large V. Singh$^1$}

\affiliation{$^1$Department of Physics, Banaras Hindu University, Varanasi 221005, India}
\affiliation{$^2$Nuclear Physics Division, Bhabha Atomic Research Center, Mumbai 400085, India}
\affiliation{$^3$Homi Bhabha National Institute, Anushaktinagar, Mumbai 400094, India}


\date{\today}

\begin{abstract}
  The meson spectra provide insight into the particle production mechanism and interaction 
in the hadronic and quark gluon plasma (QGP) phases.  
  The detailed study of systematics of meson spectra is important also because it acts as 
ingredient for estimating the hadronic decay backgrounds in the photon, single lepton and 
dilepton spectra which are the penetrating probes of quark gluon plasma.
 In this work, we parameterize experimentally measured pion spectra 
and then obtain the spectra of other light mesons using a property known as $m_T$ scaling. 
  The $m_{T}$  scaled spectra for each meson is compared with experimental data for
p+p, d+Au and Au+Au systems at $\surd s_{NN} $ = 200 GeV.  The agreement of the  
$m_{T}$ scaled and experimental data shapes are excellent in most cases and their 
fitted relative normalization gives ratio of meson to pion $m_T$ spectra. 
These ratios are useful to obtain the hadronic decay contribution in 
photonic and leptonic channels but also point to the quantitative changes in the dynamics of the 
heavy ion collision over p+p collisions. It is shown that, the particles with charm contents 
behave differently as compared to pions in d+Au systems and particles either with strange or
charm contents behave differently from pions in Au+Au systems. For Au+Au system, three 
centrality classes have been studied which reveal that for the particles like kaon and $\phi$,
peripheral collision data is better reproduced as compared to central and their relative 
ratios with pions also increase as the collisions become more central.  
\end{abstract}



\maketitle

\section{Introduction}

 Heavy ion collisions at relativistic energies are performed to create and study dense and/or
hot matter in the laboratory. 
  In Au+Au collisions at $\sqrt s_{NN}$ = 200 GeV at RHIC, many signals point to the formation 
of Quark Gluon Plasma (QGP) \cite{WP}. At RHIC, currently detailed properties of the strongly 
interacting matter are under investigation using a variety of observables. 
  To isolate phenomena related to the dense and hot medium 
created in such collisions and to understand cold nuclear matter effects, it is also important 
to measure particle production in smaller collision systems like p+p and d+A. 
   Measurements of transverse momentum spectra for particles emerging from p+p collisions 
are used as a baseline to which similar measurements from heavy ion collisions are compared. 
In addition, several observations from p+p collisions, such as the $p_T$ spectra with particle
mass are interesting in their own right. 
  The nuclear modification factor $R_{AA}$ for several identified hadrons at high transfer momentum serve 
as a probe for jet quenching \cite{RAA}. It has been observed that there is strong suppression of 
hadrons in Au+Au collisions while no suppression is observed in d+Au collisions \cite{HADPHEN}. 
  The photons $R_{AA}$ remains flat both for d+Au as well as for Au+Au collisions even at 
high $p_T$ \cite{PHOPHEN}. 
  Being electromagnetic the photons escape unaffected and their spectra after subtracting the hadronic 
decay contribution reflect the properties of the medium such as temperature.
  The dielectron invariant mass measured by PHENIX collaboration shows enhancement over the cocktail
from hadronic contribution in low mass region and is also associated to the thermal radiation 
from QGP \cite{DIELEC}. 
  The single electrons coming from semi electronic 
decays of charm and beauty quarks are important hard probes of the properties of matter
produced in heavy ion collisions \cite{SINGLE}. 
 In order to obtain the electron spectra from charm 
decays one needs to subtract the electron contribution from other meson decays \cite{COCK}.  
  Getting the cocktail of single electrons, dielectrons and 
photon coming from the decays of all mesonic sources is crucial in any of the above analysis.

 WA80 collaboration found that the spectral shapes of $\pi$ and $\eta$ mesons are identical
when plotted as a function of $m_T$ \cite{WA80}. This property is known as $m_T$ scaling
and has been extremely useful to obtain the unknown meson spectra.
 
 In this work, we used $m_{T}$ scaling to obtain all 
mesonic spectra from a given meson spectrum. 
  We parameterize pion spectra first and 
then we obtain the spectra of other light mesons using $m_T$ scaling. 
The relative normalization of the  $m_{T}$  scaled spectra is then fitted to the experimental 
data for all mesons in p+p, d+Au and Au+Au collisions.
 Both the magnitudes and shapes of the  $m_{T}$  scaled and experimental data 
are studied for all these systems for many particles namely   
K, $\eta$, $\phi$, $J/\psi$ and $\omega$ mesons.

\section{Fit procedure using $m_{T}$ scaling}

In this section, we describe the fitting procedure using $m_{T}$ scaling, but before that 
we give a brief theoretical background of the fit function used in our analysis. 
  Hagedorn proposed the following imperical formula \cite{HAG1} to describe
the data of invariant cross section of hadrons as a function of $p_{T}$ over a wide
range (0.3-10 GeV/$c$).
Hagedorn described this as "inspired by QCD".

\begin{eqnarray}
 E{d^3N \over dp^3}  =  \frac {A} {(1 + {p_{T} \over p_0})^n }.
\end{eqnarray}
Here $A$, $p_0$ and $n$ are fit parameters.
The two limiting cases of this formula are as follows: 
 
\begin{eqnarray}
  \frac {1} {(1 + {p_{T}\over p_0})^n }   & \simeq &  {\rm exp}\left(\frac {-np_{T}} {p_0}\right), \, \, \, \,\,  {\rm for}\,\,\, p_{T} \rightarrow 0 \\
  & \simeq  &\left(\frac {p_0} {p_{T}}\right)^n,  \,\,\,\,\,\,{\rm for}\,\,\, p_{T} \rightarrow \infty. 
\end{eqnarray}  
  At low transverse momenta it assumes an exponential form and at large transverse momenta 
it becomes a power law arises from "QCD inspired" quark interchange
model \cite{HAGEFACT} as:

\begin{eqnarray}
  E{d^3N \over dp^3} \sim  (m_{T}^2)^{-4} \sim \frac {1} {(p_{T})^8} 
\end{eqnarray}  

  Even though the inclusive cross section is dominated by low $p_{T}$ particles, the hardening of the 
$p_{T}$ distribution with increasing center of mass energy implies 
an increase of the transverse momentum. To better describe both low as well as high $p_{T}$ range, 
UA1 collaboration \cite{UA1} used a hybrid form as follows:

\begin{eqnarray}
  E{d^3N \over dp^3} & = & B \, {\rm exp}(-b m_{T}),\, \, \,  {\rm for}\,\, p_{T} < p_{T}^* \\
  &  =  & \frac {A} {(1 + {p_{T} \over p_0})^n }, \, \, \, \, \,   {\rm for} \,\, p_{T} > p_{T}^*. 
\end{eqnarray} 
with $p_{T}^*$ as a free parameter.
 The PHENIX collaboration obtained a single form referred as 
modified Hagedorn formula \cite{DIELEC, PPG077, PPG099}.
 The modification is to better 
describe the $\pi^0$ spectrum for wider $p_{T}$ range, in particular at high $p_{T}$ 
where the spectrum behaves close  to a simple power law function.
 This single formula has been used to successfully describe the hadron spectra measured 
in p+p \cite{DIELEC, PPG077, PPG099, SEANA}, d+Au \cite{SEANA} as well as in Au+Au \cite{DIELEC, PPG077, SEANA} 
collisions at different energies and is given by 
  
\begin{eqnarray}
  E{d^3N \over dp^3}  =  \frac {A} { \left[ {\rm exp}(- a p_{T} - b p_{T}^2) 
      + {p_{T} \over p_0} \right]^n }   
\end{eqnarray}
which is close to a Hagedorn function at low $p_{T}$ and satisfies the requirement that
the function is a pure power law at high $p_{T}$ .
 The term quadratic in $p_{T}$ in the exponential term in the denominator is more important for 
the Au+Au collisions.  

  In our analysis, first we parameterize experimentally measured pion spectra.
 In the fitting procedure, both neutral as well as charged pion data are incorporated.
This is based on the assumption that the neutral pion spectrum is the same as the 
average charged pion spectrum. The fit function used here is the same 
modified Hagedorn distribution used by PHENIX collaboration but we replace $p_T$ by 
$m_T$ written as 

\begin{eqnarray}
 E{d^3N \over dp^3} & = & \frac {A} { \left[ {\rm exp}(- a m_{T} - b m_{T}^2) 
                             + {m_{T} \over p_0} \right]^n },  \nonumber \\ 
      & =  & f_{\pi} \left( \sqrt{p_T^2 + m_{\pi}^2} \right),
\end{eqnarray}
where  $A$, $a$, $b$, $p_0$ and  $n$ are the fit parameters. The pion spectra measured in 
p+p, d+Au and Au+Au systems are fitted using this distribution and the 
parameters obtained are given in table \ref{pppara}. For Au+Au system, data corresponding
to three centrality classes namely 0-20 \%, 20-60 \% and 60-92 \% has been analyzed.
The fit parameters for the centrality data for pions are given in table \ref{auaucentPar}. 
 Comparing the values of $b$ for different systems shows that 
term quadratic in $m_{T}$ in the exponential term in the denominator is more important for 
the Au+Au collisions. We notice that the power $n$ for all systems is close to 8 providing 
qualitative theoretical support for the quark interchange model.

  The light mesons which contribute sizeably to any measured electron and/or photons via their decay
are pions, $\eta$, $\rho$, $\omega$, $\phi$, $\eta\prime$ etc. The $\eta$ meson contributes a sizeable 
fraction of decay electrons, in particular at high $p_{T}$. Using $m_{T}$ scaling, 
we  obtain the spectra of these light neutral  mesons using pion fit function as 

\begin{eqnarray}
 E{d^3N \over dp^3}  =  S \,\, f_{\pi} \left( \sqrt{p_T^2 + m_{h}^2} \right)
\end{eqnarray}
where $m_{h}$ is the rest mass of the corresponding 
hadron or meson. The factor $S$ is the relative normalization of the meson $m_T$ spectrum
to the pion $m_T$ spectrum which we obtain by fitting the experimentally measured meson spectrum.

The factor $S$ should be close to, but not identical to, some
 standard meson/$\pi^0$ ratios in the literature \cite{PPG065} that have been
 obtained as averaged over $p_{T}$ intervals:

$\eta$/$\pi^0$=0.48 $\pm$ 0.03 \cite{PPG055};

$\rho$/$\pi^0$=1.0 $\pm$ 0.3, predicted by \cite{PYTHIA}; 

$\omega$/$\pi^0$=0.9 $\pm$ 0.06 \cite{RATOMEGA};

$\eta\prime$/$\pi^0$=0.40 $\pm$ 0.12, predicted by PYTHIA \cite{PYTHIA};

$\phi$/$\pi^0$= 0.25 $\pm$ 0.08 \cite{VICTOR};

These values are obtained mostly by p+p measurements and models for p+p collisions,
but also are widely used in cocktail calculations for the d+Au system.
 For Au+Au system the ratios are given in Ref.\cite{RATPHI}. 
 All the meson data used in this analysis along with mode of measurement and 
$p_{T}$ ranges with their references are listed  
in table \ref{refer1} for p+p and in \ref{refer2} for d+Au and Au+Au systems. 
The particles like pions and kaons are measured by 
time of flight (TOF). The errors on the data are quadratic sums of statistical 
and uncorrelated systematic errors wherever available. 
  All the data used in the present work is for same rapidity ( $|y| < 0.35 $ )
and from PNENIX experiment at RHIC. 

\begin{table}
\caption{The parameters of the Hagedorn distribution obtained by fitting pion spectra 
measured in p+p, d+Au and Au+Au collisions at $\sqrt{s_{NN}}$ = 200 GeV.}
\label{pppara}
\begin{tabular}{|c|c|c|c|}
\hline
                 &  \multicolumn{3}{|c|} {Collision systems}              \\
\cline{2-4}
 Parameters      &    \rm p+p               &  \rm d+Au              &   \rm Au+Au         \\
\hline           
$A$ (GeV/$c$)$^{-2}$    & 11.11  $\pm$ 0.37     & 52.30 $\pm$ 1.63      & 822.30 $\pm$ 15.90     \\
$a$ (GeV/$c$)$^{-1}$  & 0.32   $\pm$ 0.03     & 0.24  $\pm$ 0.01      & 0.420 $\pm$ 0.006     \\
$b$ (GeV/$c$)$^{-2}$    & 0.024  $\pm$ 0.011    & 0.11  $\pm$ 0.01      & 0.215 $\pm$ 0.006     \\
$p_0$ (GeV/$c$)      & 0.72   $\pm$ 0.03     & 0.77  $\pm$ 0.02      & 0.697 $\pm$ 0.003     \\
$n$                  & 8.42   $\pm$ 0.12     & 8.46  $\pm$ 0.07      & 8.35 $\pm$ 0.01     \\
\hline
\end{tabular}
\end{table}

\begin{table}
\caption{The parameters of the Hagedorn distribution obtained by fitting pion spectra 
measured in Au+Au collisions of different centralities at $\sqrt{s_{NN}}$ = 200 GeV.}
\label{auaucentPar}
\begin{tabular}{|c|c|c|c|}
\hline
                      &  \multicolumn{3}{|c|} {Au+Au collision centrality}              \\
\cline{2-4}
 Parameters           &    0-20 \%               &  20-60 \%              &   60-92 \%             \\
\hline           
$A$ (GeV/$c$)$^{-2}$   & 1883.23  $\pm$ 31.20     & 744.77 $\pm$ 9.11      & 132.99  $\pm$ 2.21     \\
$a$ (GeV/$c$)$^{-1}$   & 0.442    $\pm$ 0.006     & 0.385  $\pm$ 0.004     & 0.273   $\pm$ 0.005     \\
$b$ (GeV/$c$)$^{-2}$   & 0.242    $\pm$ 0.006     & 0.192  $\pm$ 0.004     & 0.154   $\pm$ 0.006     \\
$p_0$ (GeV/$c$)       & 0.708    $\pm$ 0.003     & 0.694  $\pm$ 0.003     & 0.653   $\pm$ 0.006     \\
$n$                   & 8.40     $\pm$ 0.01      & 8.26   $\pm$ 0.01      & 8.14    $\pm$ 0.03     \\
\hline
\end{tabular}
\end{table}

\begin{table}
\caption{Particles with their measured decay channels and $p_{T}$ range for different p+p collisions with references. The data is for central rapidity region $|y|<0.35$.}
\label{refer1}
\begin{tabular}{|c|c|c|c|}
\hline
\cline{1-4}
  Particle    &    Mode                   & $p_T$ range        &  Reference              \\
\hline
$\pi^{0}$        & $ \gamma \gamma$          & 0.6-18.9 GeV   &  \cite{PPG063}          \\
$\pi^{\pm}$      &    TOF                     & 0.3-2.6 GeV    &  \cite{PPG030}          \\
$K^{\pm}$        &    TOF                     & 0.4-1.8 GeV    &    \cite{PPG030}        \\
$\eta$         & $ \gamma \gamma$          &  2.7-11.0 GeV   &  \cite{PPG055}           \\
               &                            &   5.5-21 GeV      &    \cite{PPG115}        \\
$\phi$         & $ e^{+} e^{-}$               &  0-3.5 GeV        &  \cite{PPG099}       \\ 
               & $K^{+} K^{-}$                &  1-7.0 GeV        &  \cite{PPG096}        \\ 
$\omega$       & $e^{+}  e^{-}$               &  0.1-3.5 GeV     & \cite{PPG099}         \\
               & $\pi^{+}  \pi^{-}$           &  2.5-9.3 GeV     & \cite{PPG064}          \\ 
               & $\pi^{0} \pi^{+}  \pi^{-}$    &  2.25-13 GeV     & \cite{PPG099}        \\ 
J/$\psi$       & $e^{+} e^{-}$                &  0.1-8.5 GeV       &  \cite{PPG069}      \\
               &                            &  0.1-8.5 GeV       &  \cite{PPG097}      \\
\hline
\end{tabular}
\end{table}

\begin{table}
\caption{Particles with their measured decay channels and $p_{T}$ range for d+Au and Au+Au systems with references. 
The data is for central rapidity region $|y|<0.35$.}
\label{refer2}
\begin{tabular}{|c|c|c|c|}
\hline
\cline{1-4}
  Particle    &    Mode                   & $p_T$ range        &  Reference              \\
\hline
   \multicolumn{4}{|c|} {d+Au collision}                           \\
\hline

$\pi^{0}$      &   $\gamma \gamma$        & 1.2-17.0 GeV     &   \cite{PPG044}      \\
$\pi^{\pm}$    &    TOF                    &  0.3-2.6 GeV      &  \cite{PPG030}         \\
$K^{\pm}$      &    TOF                    &  0.45-1.8 GeV     &   \cite{PPG030}       \\
$\eta$       &  $\gamma \gamma$          &  1.2-4.75 GeV     &  \cite{PPG055}       \\
             &                           &  2.25-11.0 GeV    &   \cite{PPG044}     \\
$\phi$       &  $K^{+} K^{-}$             &  1.1-7 GeV         &  \cite{PPG096}         \\ 
$\omega$     &  $\pi^{0} \gamma$         &  3-9 GeV           &   \cite{PPG064}       \\ 
J/$\psi$     & $e^{+} e^{-}$              &  0.5-4.5 GeV       &  \cite{PPG038}       \\
\hline
   \multicolumn{4}{|c|} {Au +Au collision}                                         \\
\hline
$\pi^{0}$     &  $\gamma \gamma$         & 1.2-19.0 GeV     &   \cite{PPG080}             \\
$\pi^{\pm}$   &      TOF                  & 0.25-2.95 GeV     &   \cite{PPG026}            \\
$K^{\pm}$     &      TOF                  & 0.45-1.95 GeV    &  \cite{PPG026}             \\
$\eta$       &  $\gamma \gamma$         &  2.25-9.5 GeV   & \cite{PPG051}          \\
             &                          &  5.5-21 GeV     & \cite{PPG115}         \\
$\phi$       &  $K^{+} K^{-}$            &  1.1-7 GeV       & \cite{PPG096}         \\ 
$\omega$     &  $\pi^{0} \gamma$         &  4.5-8.5 GeV    &  \cite{PPG118}       \\ 
J/$\psi$     &  $ e^{+} e^{-}$           &  0.25-5 GeV      & \cite{PPG068}          \\
\hline
\end{tabular}
\end{table}

\section{Results}

\begin{figure}
\includegraphics[width=0.9\textwidth]{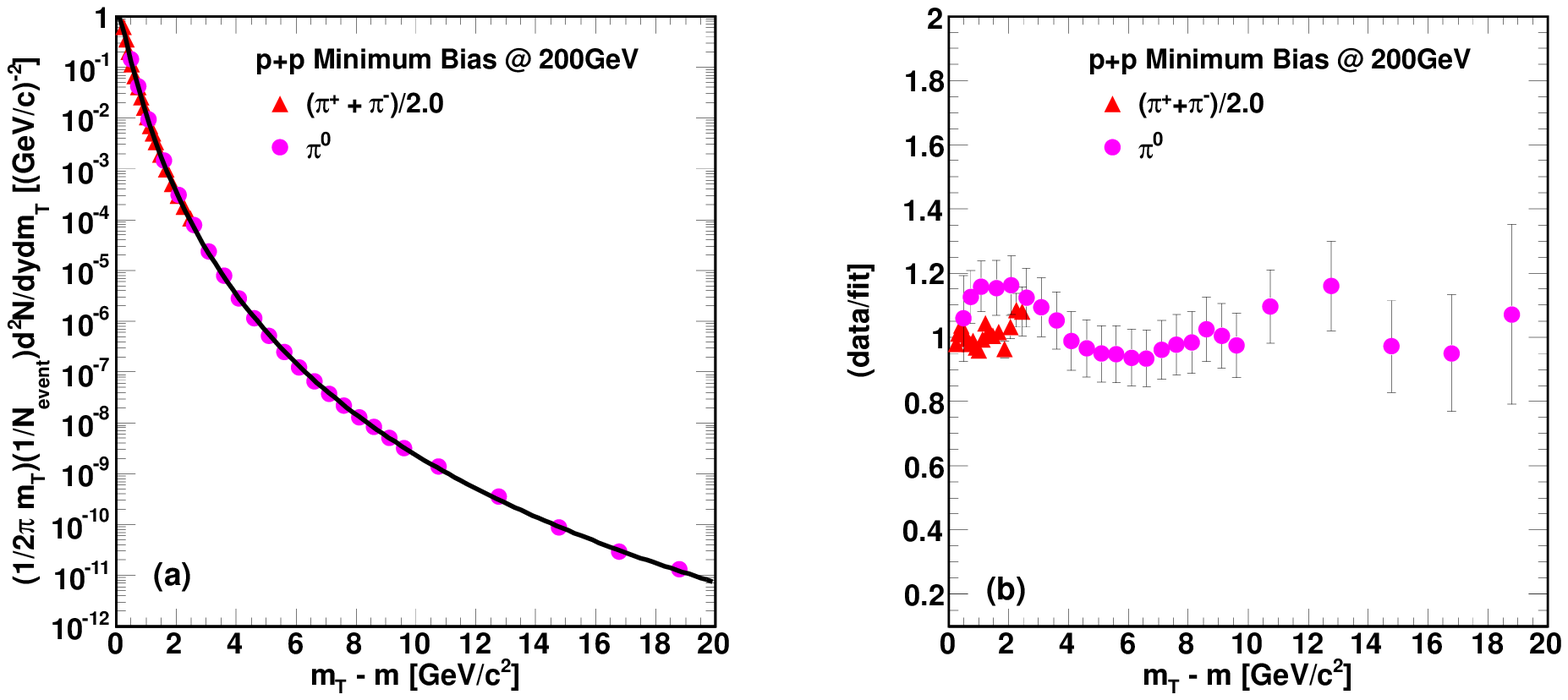}
\caption{(Color online) a) the invariant yields of neutral \cite{PPG063} and charged 
pions \cite{PPG030} as a function of $m_{T}$ measured in p+p collision at 200 GeV fitted 
with the Hagedorn function. b) the ratio of data to the fit.}
\label{pppion}
\end{figure}
\begin{figure}
\includegraphics[width=0.9\textwidth]{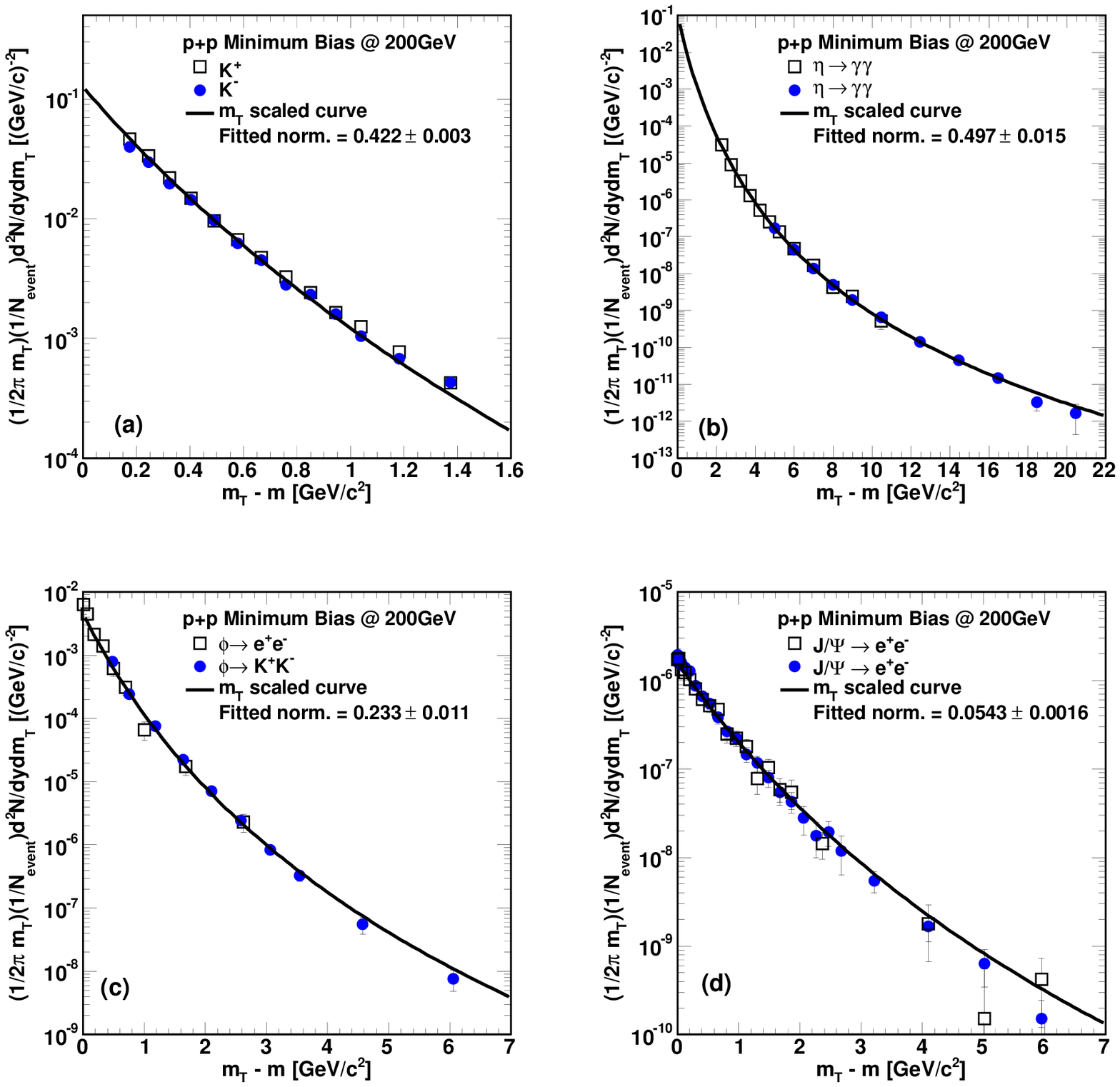}
\caption{(Color online) The invariant yield of K$^\pm$ \cite{PPG030}, 
$\eta$ (open squares \cite{PPG055}, solid circles \cite{PPG115}), 
$\phi$ (open squares \cite{PPG099}, solid circles \cite{PPG096})
and $J/\psi$ (open squares \cite{PPG069}, solid circles \cite{PPG097}) as a 
function of $m_{T}$ measured in p+p system. The solid line is obtained using $m_{T}$ scaling; 
the relative normalization has been used to fit the measured spectra.}
\label{ppmesons}
\end{figure}
\begin{figure}
\includegraphics[width=0.5\textwidth]{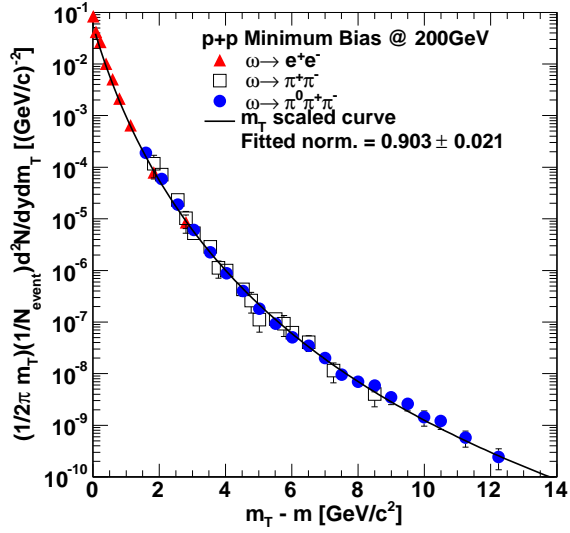}
\caption{(Color online) The invariant yield of $\omega$ meson (solid triangles \cite{PPG099}, 
open squares \cite{PPG064}, solid circles \cite{PPG099})  
as a function of $m_{T}$ measured in p+p system along with the $m_{T}$ scaled 
curve (solid line).}
\label{ppomega}
\end{figure}

  Figure (\ref{pppion}a) shows the invariant yields of neutral 
\cite{PPG063} and charged pions \cite{PPG030} as a function of $m_{T}$ measured 
in p+p collision at $\sqrt{s_{NN}}$ = 200 GeV fitted with the 
Hagedorn function. The figure (\ref{pppion}b) shows the ratio of data to the fit.
  Figure (\ref{ppmesons}) shows the invariant yields of K$^\pm$ \cite{PPG030}, 
$\eta$ \cite{PPG115,PPG055}, $\phi$ \cite{PPG096,PPG099} and $J/\psi$ \cite{PPG069,PPG097} 
as a function of $m_{T}$ measured in p+p system. The solid line is obtained using 
$m_{T}$ scaling; the relative normalization has been used to fit the measured spectra. 
 Figure (\ref{ppomega}) shows the invariant yield of $\omega$ meson \cite{PPG064,PPG099} 
as a function of $m_{T}$ measured in p+p system along with the $m_{T}$ scaled curve (solid line).

\begin{figure}
\includegraphics[width=0.9\textwidth]{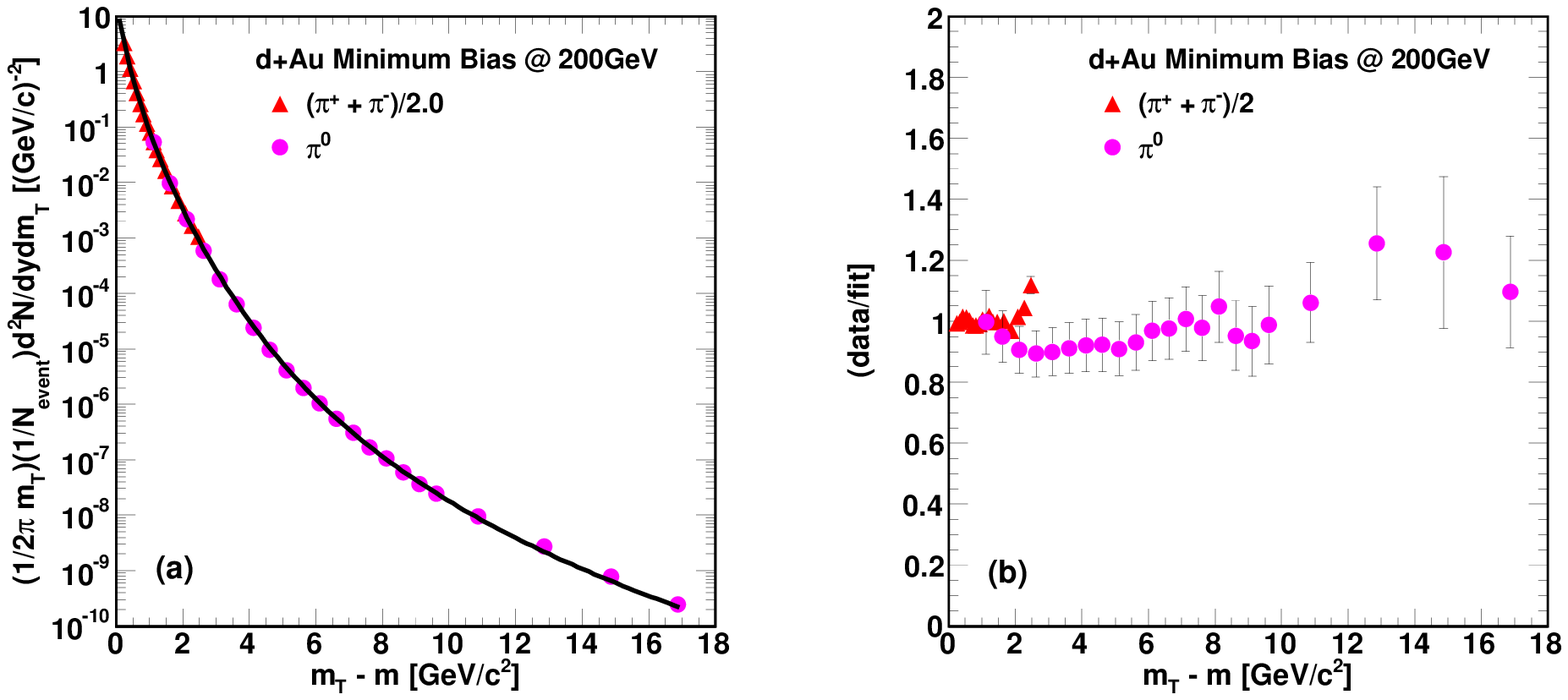}
\caption{(Color online) a) the invariant yields of neutral \cite{PPG044} and charged 
pions \cite{PPG030} as a function of $m_{T}$ measured in d+Au at 200 GeV fitted 
with the Hagedorn function. b) the ratio of data to fit.}
\label{daupion}
\end{figure}
\begin{figure}
\includegraphics[width=0.9\textwidth]{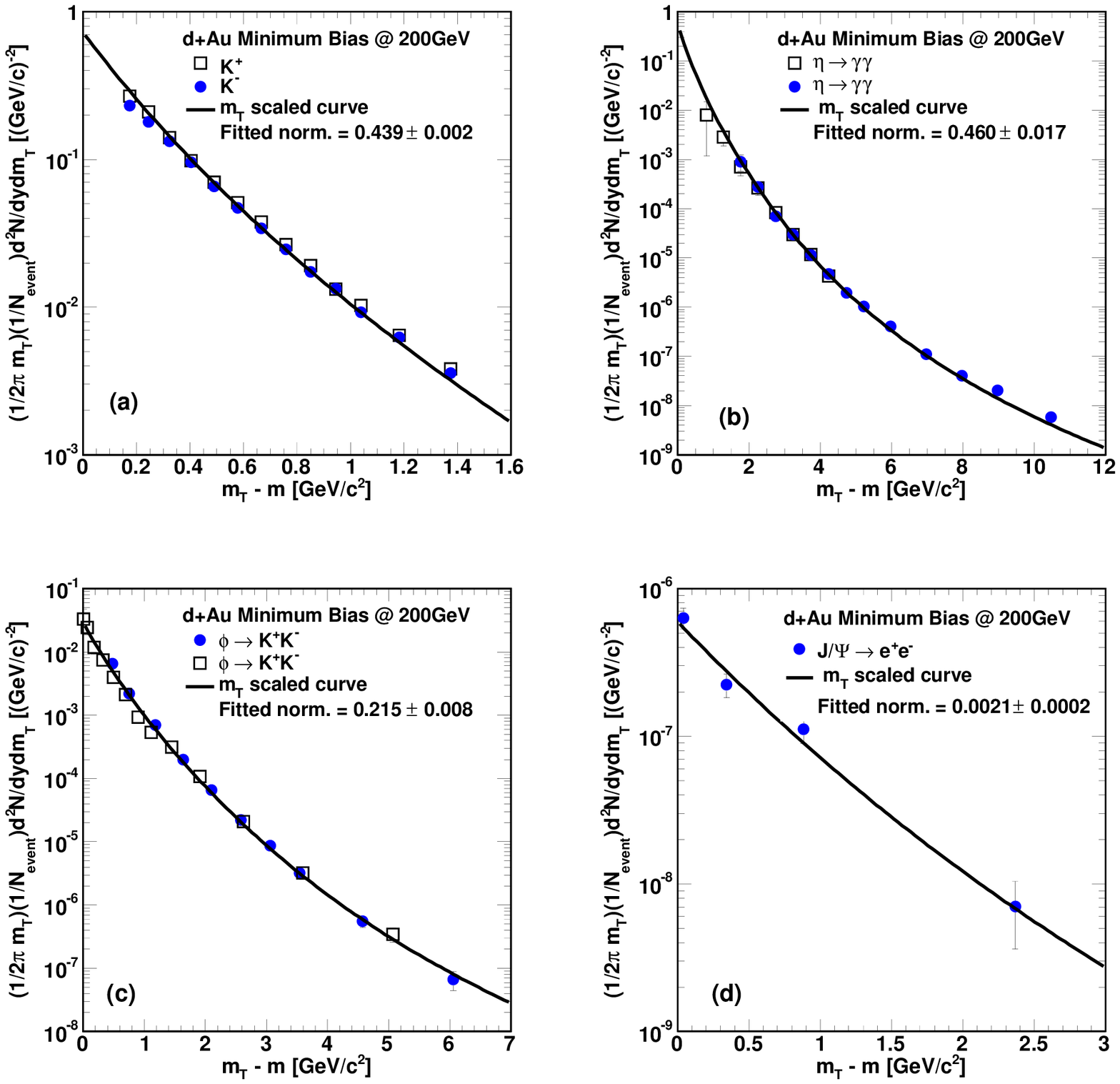}
\caption{(Color online) The invariant yield of measured K$^\pm$ \cite{PPG030}, 
$\eta$ (open squares \cite{PPG055}, solid circles \cite{PPG044}), 
$\phi$ \cite{PPG096} and $J/\psi$ \cite{PPG038} as a function of $m_{T}$ measured in d+Au 
system. The solid line is obtained using $m_{T}$ scaling; 
the relative normalization has been used to fit the measured spectra.}
\label{daumesons}
\end{figure}
\begin{figure}
\includegraphics[width=0.5\textwidth]{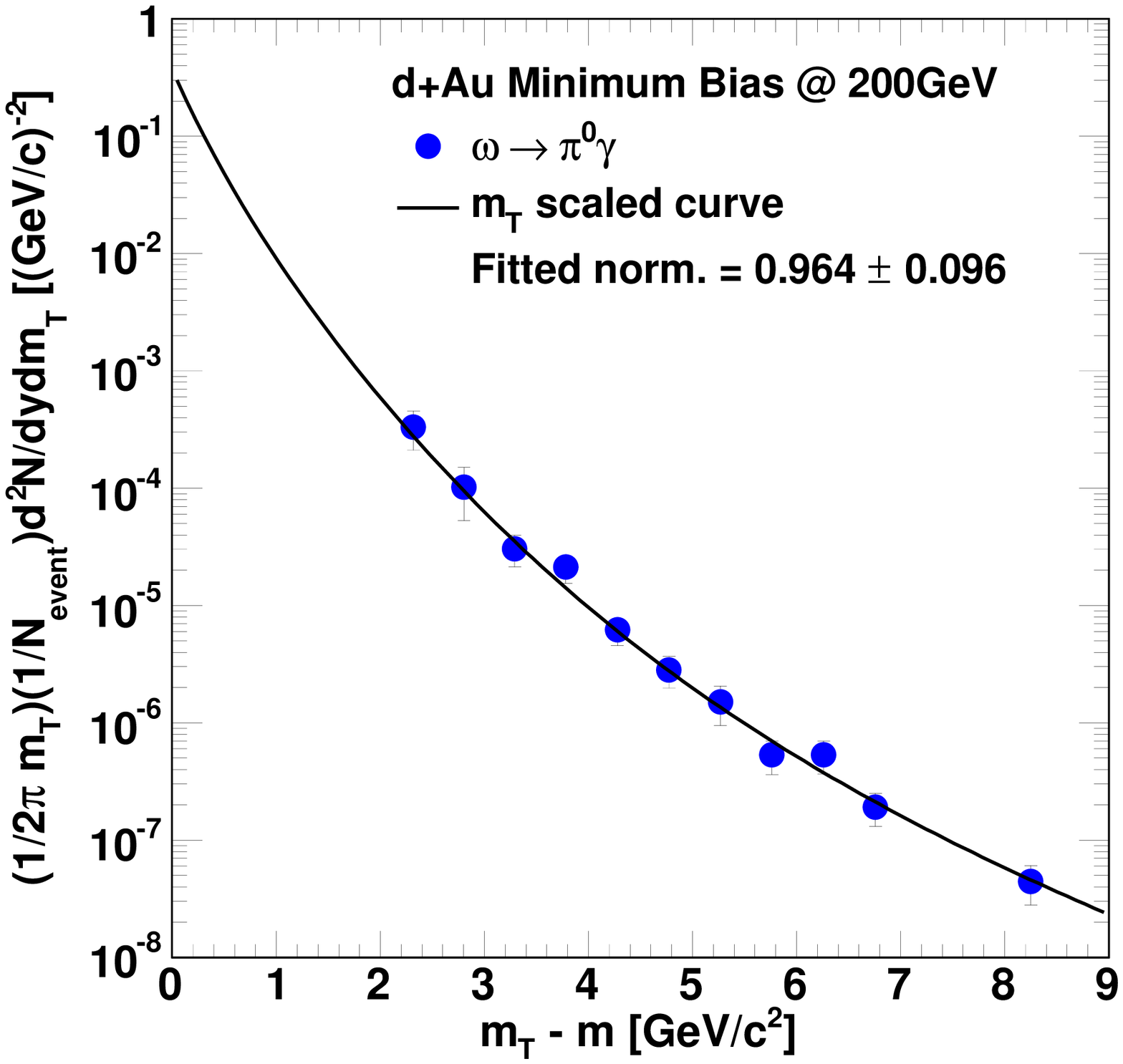}
\caption{(Color online) The invariant yield of $\omega$ \cite{PPG064} meson as a 
function of $m_{T}$ measured in d+Au system along with $m_{T}$ scaled curve.}
\label{dauomega}
\end{figure}

 Figure (\ref{daupion}a) shows the invariant yields of neutral \cite{PPG044} and 
charged pions \cite{PPG030} as a function of $m_{T}$ measured in d+Au at 
$\sqrt{s_{NN}}$ = 200 GeV fitted with the 
Hagedorn function. The figure (\ref{daupion}b) shows the ratio of data to fit.
  Figure (\ref{daumesons}) shows the invariant yields of measured K$^\pm$ \cite{PPG030}, 
$\eta$ \cite{PPG044,PPG055}, $\phi$ \cite{PPG096} and $J/\psi$ \cite{PPG038} as a 
function of $m_{T}$ measured in d+Au system. The solid line is obtained using $m_{T}$ scaling; 
the relative normalization has been used to fit the measured spectra. 
 Figure (\ref{dauomega}) shows the invariant yield of $\omega$ meson \cite{PPG064} as a 
function of $m_{T}$ measured in d+Au system along with $m_{T}$ scaled curve.

\begin{figure}
\includegraphics[width=0.9\textwidth]{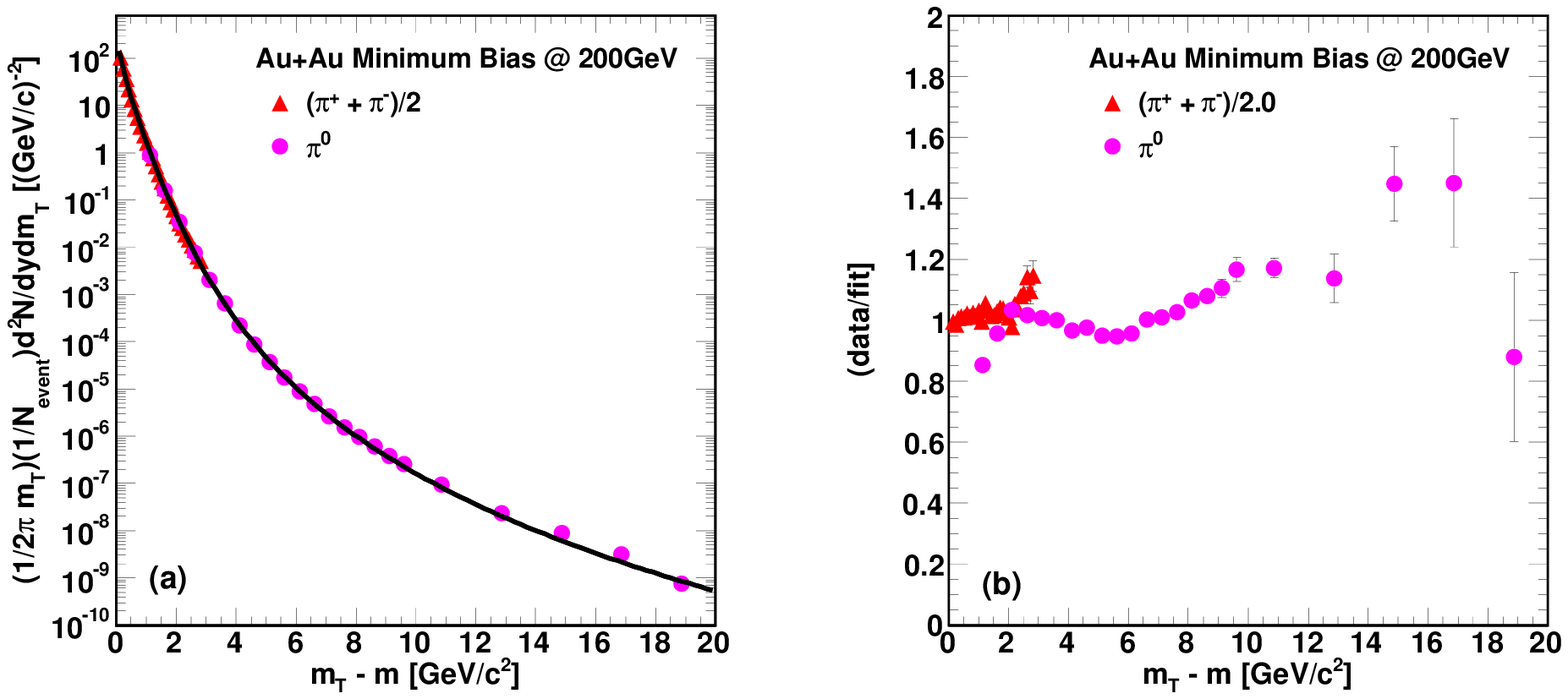}
\caption{(Color online) a) the invariant yields of neutral \cite{PPG080} and charged pions \cite{PPG026}  
as a function of $m_{T}$ measured in Au+Au at 200 GeV fitted with 
the Hagedorn function. b) the ratio of data to fit.}
\label{auaupion}
\end{figure}
\begin{figure}
\includegraphics[width=0.9\textwidth]{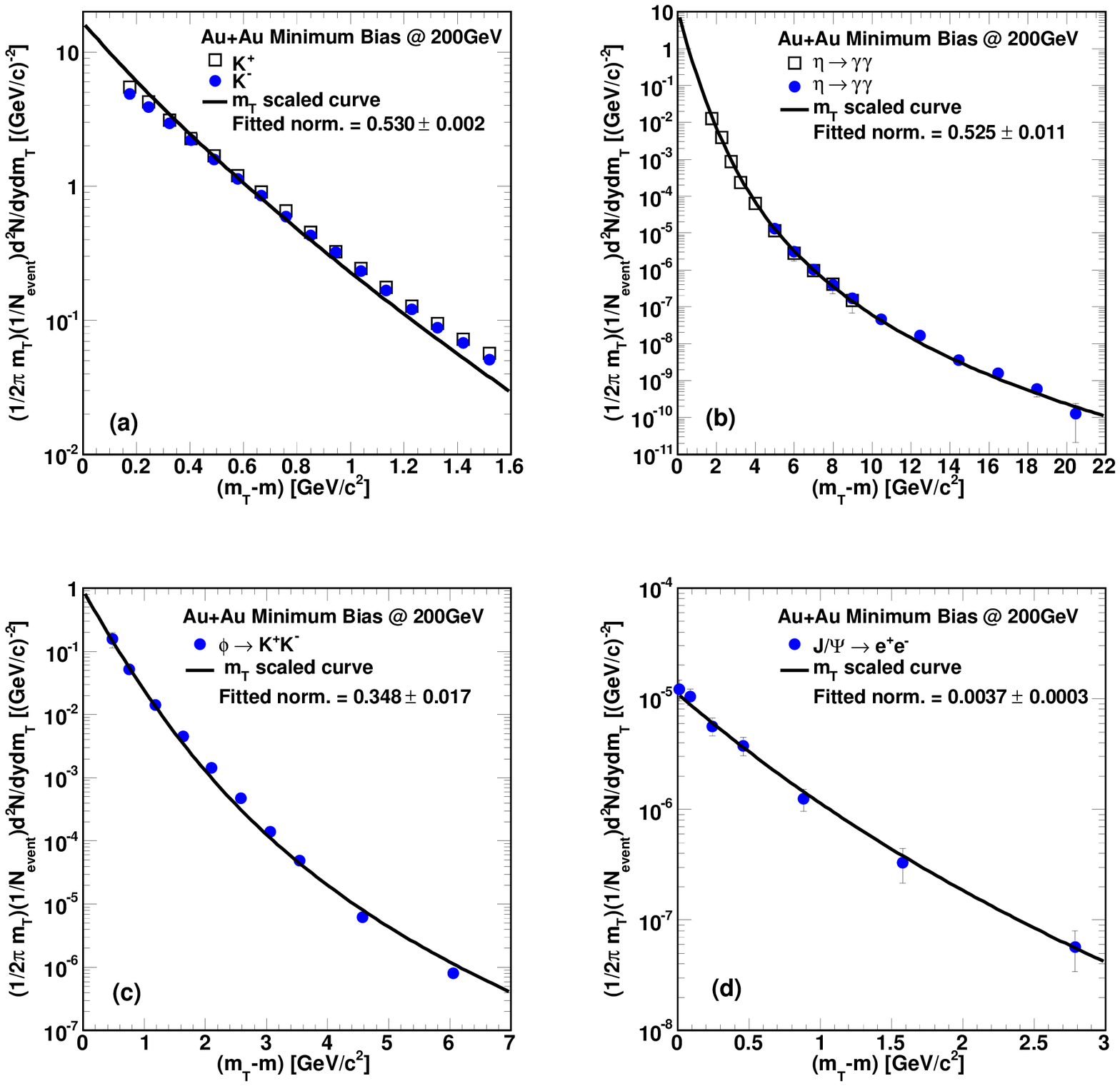}
\caption{(Color online) The invariant yield of measured K$^\pm$ \cite{PPG026}, 
$\eta$ (open squares \cite{PPG051}, solid circles \cite{PPG115}), 
$\phi$ \cite{PPG096}, $J/\psi$ \cite{PPG068} as a function 
of $m_{T}$ in Au+Au system. The solid line is obtained using $m_{T}$ scaling;
the relative normalization has been used to fit the measured spectra. }
\label{auaumesons}
\end{figure}

 Figure (\ref{auaupion}a) shows the invariant yields of neutral \cite{PPG080} 
and charged pions \cite{PPG026} as a function of $m_{T}$ measured in Au+Au at 
$\sqrt{s_{NN}}$ = 200 GeV fitted with 
the Hagedorn function. The figure (\ref{auaupion}b) shows the ratio of data to fit.
  Figure (\ref{auaumesons}) shows the invariant yields of measured K$^\pm$ \cite{PPG026}, 
$\eta$ \cite{PPG115,PPG051}, $\phi$ \cite{PPG096} and $J/\psi$ \cite{PPG068} as a function 
of $m_{T}$ in Au+Au system. The solid line is obtained using $m_{T}$ scaling;
the relative normalization has been used to fit the measured spectra. 

\begin{figure}
\includegraphics[width=0.5\textwidth]{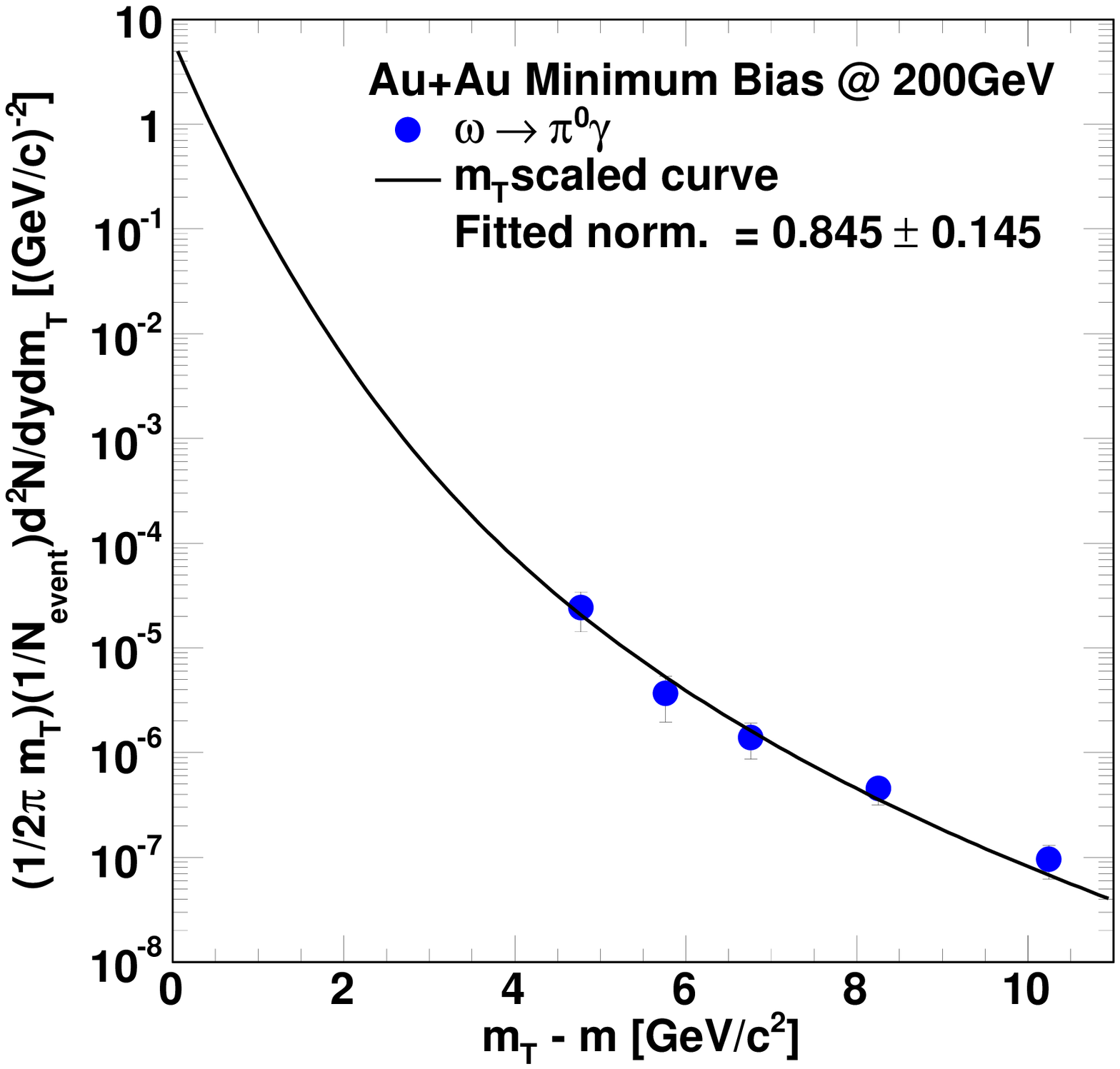}
\caption{(Color online) The invariant yield of $\omega$ \cite{PPG118} meson as a 
function of $m_{T}$ measured in Au+Au system along with $m_{T}$ scaled curve.}
\label{auauomega}
\end{figure}

 Figure (\ref{auauomega}) shows the 
invariant yield of $\omega$ \cite{PPG118} meson as a 
function of $m_{T}$ measured in Au+Au system along with $m_{T}$ scaled curve.

\begin{table}
\caption{The relative normalization (meson/$\pi^0$) obtained by fitting the $m_T$ scaled spectra 
with the measured spectra for different collision systems.}
\label{ratios}
\begin{tabular}{|c|c|c|c|}
\hline
                 &  \multicolumn{3}{|c|} {Collision systems}              \\
\cline{2-4}
 particle ratio $S$  &  \rm p+p            &  \rm d+Au            &   \rm Au+Au (MB)      \\
\hline
 K/$\pi^0$       & 0.422 $\pm$ 0.003   & 0.439  $\pm$ 0.002   &  0.530  $\pm$ 0.002    \\
$\eta$/$\pi^0$   & 0.497 $\pm$ 0.015   & 0.460  $\pm$ 0.017   &  0.525  $\pm$ 0.040    \\
$\phi$/$\pi^0$   & 0.233 $\pm$ 0.011   & 0.215  $\pm$ 0.008   &  0.348  $\pm$ 0.017    \\
$\omega$/$\pi^0$ & 0.903 $\pm$ 0.021   & 0.964  $\pm$ 0.10    &  0.845  $\pm$ 0.145    \\
J/$\psi$/$\pi^0$ & 0.054 $\pm$ 0.002   & 0.0021 $\pm$ 0.0002  &  0.0037 $\pm$ 0.0003   \\
\hline
\end{tabular}
\end{table}

\begin{table}
\caption{The relative normalization (meson/$\pi^0$) obtained by fitting the $m_T$ scaled spectra 
with the measured spectra for different Au+Au collision of different centralities.}
\label{ratios1}
\begin{tabular}{|c|c|c|c|c|}
\hline
                 &  \multicolumn{4}{|c|} {Au+Au collision}              \\
\cline{2-5}
 particle ratio $S$ &             MB        &  0-20\%               &  20-60\%              &  60-92\%     \\
\hline
 K/$\pi^0$       &  0.530  $\pm$ 0.002   &  0.504  $\pm$ 0.001   &  0.494  $\pm$ 0.001   &  0.486  $\pm$ 0.003  \\
$\eta$/$\pi^0$   &  0.525  $\pm$ 0.040   &  0.542  $\pm$ 0.017   &  0.580  $\pm$ 0.014   &  0.548  $\pm$ 0.024  \\
$\phi$/$\pi^0$   &  0.348  $\pm$ 0.017   &  0.401  $\pm$ 0.010   &  0.391  $\pm$ 0.007   &  0.302  $\pm$ 0.014  \\
\cline{2-5}
                 &       MB              &   0-20\%             &  20-40\%               &  40-92\% \\
\cline{2-5}
J/$\psi$/$\pi^0$ &  0.0037 $\pm$ 0.0003  &  0.0032 $\pm$ 0.0005  &  0.0043 $\pm$ 0.0005  &  0.0033 $\pm$ 0.0006 \\
\hline
\end{tabular}
\end{table}


 The table \ref{ratios} shows the normalization factor of meson to pion $m_T$ spectra
(meson/$\pi^0$) obtained by fitting the 
$m_T$ scaled spectra with the measured spectra for different collision systems.
  In case of p+p they are in agreement with the ratios available in the literature. 
One can observe that the shapes of the derived $m_T$ scaled spectra very well 
reproduce the measured spectra.

 In case of d+Au, the fitted meson to pion ratios are in agreement with those in case 
of p+p for all mesons except for the J/$\psi$. The J/$\psi$ to pion ratio is small in 
d+Au case because the J/$\psi$ yields are suppressed but the pion yields remain unaffected.
The shapes of the derived $m_T$ scaled spectra reproduce all measured spectra including 
J/$\psi$. It shows that the J/$\psi$ yields are suppressed uniformly in all the $p_T$ range 
considered here. 

 In case of Au+Au, the fitted $\eta$ to pion ratio is similar to that in case of pp. It is
because $\eta$ and pions are suppressed by similar amount. Also the shape of the curve matches well 
with the data. The same is true with $\omega$. It means that one can obtain the ratios of the 
$\eta$ and $\omega$ $m_T$ spectra with pion $m_T$ spectra for p+p collisions and then can
use them for the case of d+Au and Au+Au systems.
The kaon to pion ratio is larger than that in p+p and the $m_T$ scaled curve
does not reproduce the shape of the measured spectra that well. The $\phi$ to pion ratio is larger 
than that in p+p because the $\phi$ is not suppressed as much as pions. One can observe
in case of $\phi$ that the shape of measured spectra is not well reproduced in intermediate $m_T$ 
region of $m_{T}-m$ $\sim$ 2 to 4 GeV.. 
 This shows that the particles with strangeness contents behave differently in Au+Au case.

  The J/$\psi$ to pion ratio is small in 
Au+Au case but even smaller in d+Au. This is quite interesting and is because
the yield of pions is suppressed only in Au+Au but the yield of 
J/$\psi$ is suppressed in both systems.
  Interestingly, the shape of the derived $m_T$ scaled spectra reproduce the measured 
J/$\psi$ spectrum. It shows that the J/$\psi$ yields are suppressed uniformly in all 
the $m_T$ range considered here. 

  We have also considered three centrality classes for Au+Au system; 
0-20 \% (most central), 20-60 \% (intermediate) and 60-92 \% (peripheral).
In case of J/$\psi$ we have data available for the centralities
0-20 \%, 20-40 \%  and 40-92 \%. The class 40-92 \% in this case can be called as 
semiperipheral. Also we do not have good centrality data for $\omega$.
 The table \ref{ratios} shows the normalization factor of meson to pion $m_T$ spectra
(meson/$\pi^0$) obtained by fitting the 
$m_T$ scaled spectra with the measured spectra for different centralities.

\begin{figure}
\includegraphics[width=0.5\textwidth]{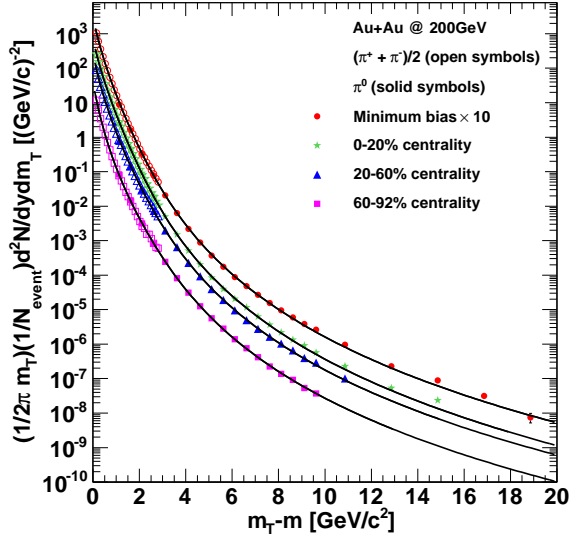}
\caption{(Color online) The invariant yield of measured 
neutral \cite{PPG080} and charged pions \cite{PPG026}
as a function of $m_{T}$ in Au+Au system for different centralities. 
The solid lines are the Hagedorn fit function.}
\label{auau10}
\end{figure}

\begin{figure}
\includegraphics[width=0.5\textwidth]{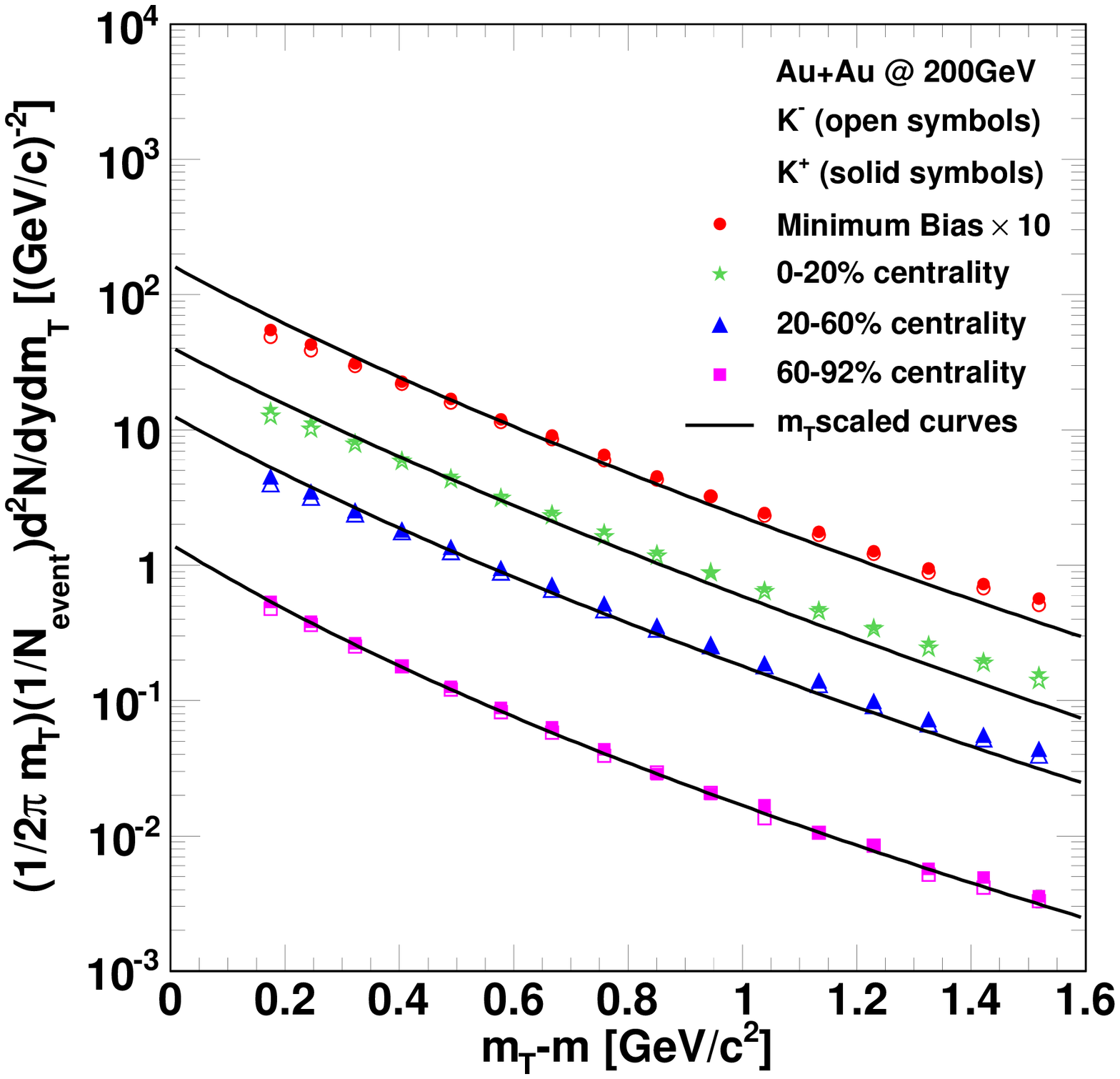}
\caption{(Color online) The invariant yield of measured K$^\pm$ \cite{PPG026} as a function 
of $m_{T}$ in Au+Au system for different centralities. The solid lines are obtained 
using $m_{T}$ scaling; the relative normalization has been used to fit the measured spectra. }
\label{auau11}
\end{figure}

\begin{figure}
\includegraphics[width=0.5\textwidth]{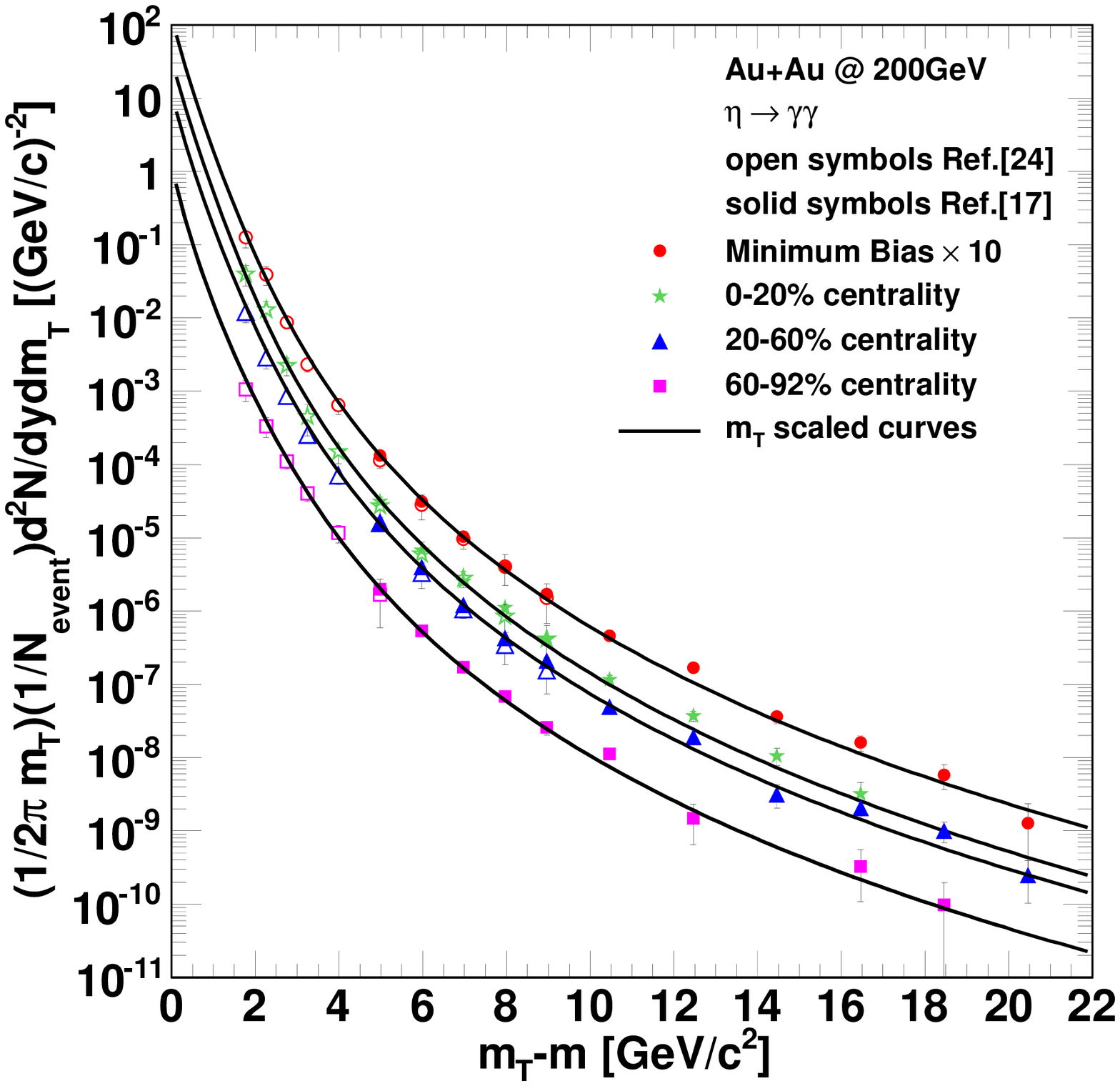}
\caption{(Color online) The invariant yield of measured 
$\eta$ (open squares \cite{PPG051}, solid circles \cite{PPG115})
as a function of $m_{T}$ in Au+Au system for different centralities. 
The solid lines are obtained using $m_{T}$ scaling;
the relative normalization has been used to fit the measured spectra. }
\label{auau12}
\end{figure}

\begin{figure}
\includegraphics[width=0.5\textwidth]{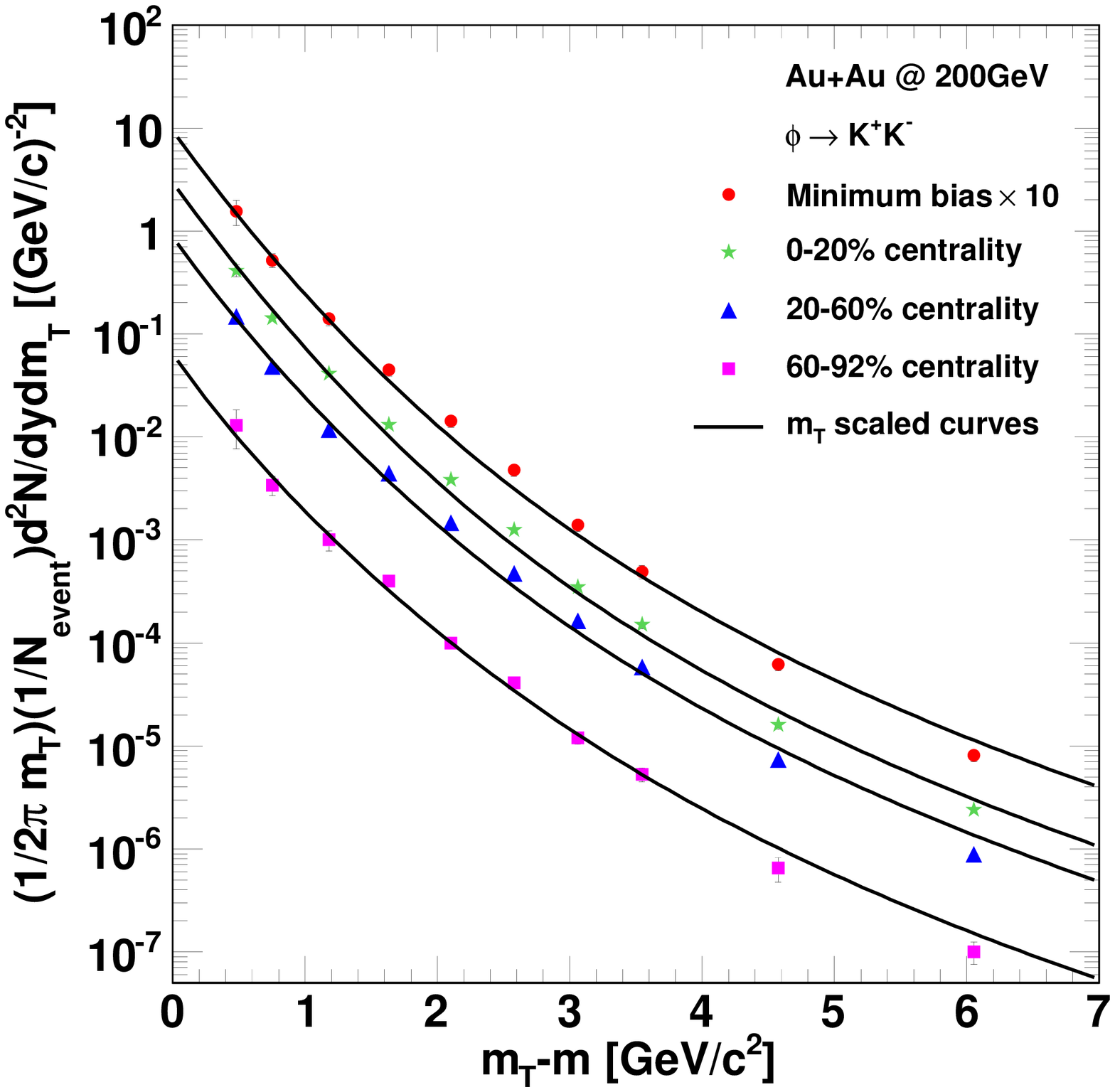}
\caption{(Color online) The invariant yield of measured $\phi$ \cite{PPG096} as a function 
of $m_{T}$ in Au+Au system for different centralities. The solid lines are obtained 
using $m_{T}$ scaling; the relative normalization has been used to fit the measured spectra. }
\label{auau13}
\end{figure}

\begin{figure}
\includegraphics[width=0.5\textwidth]{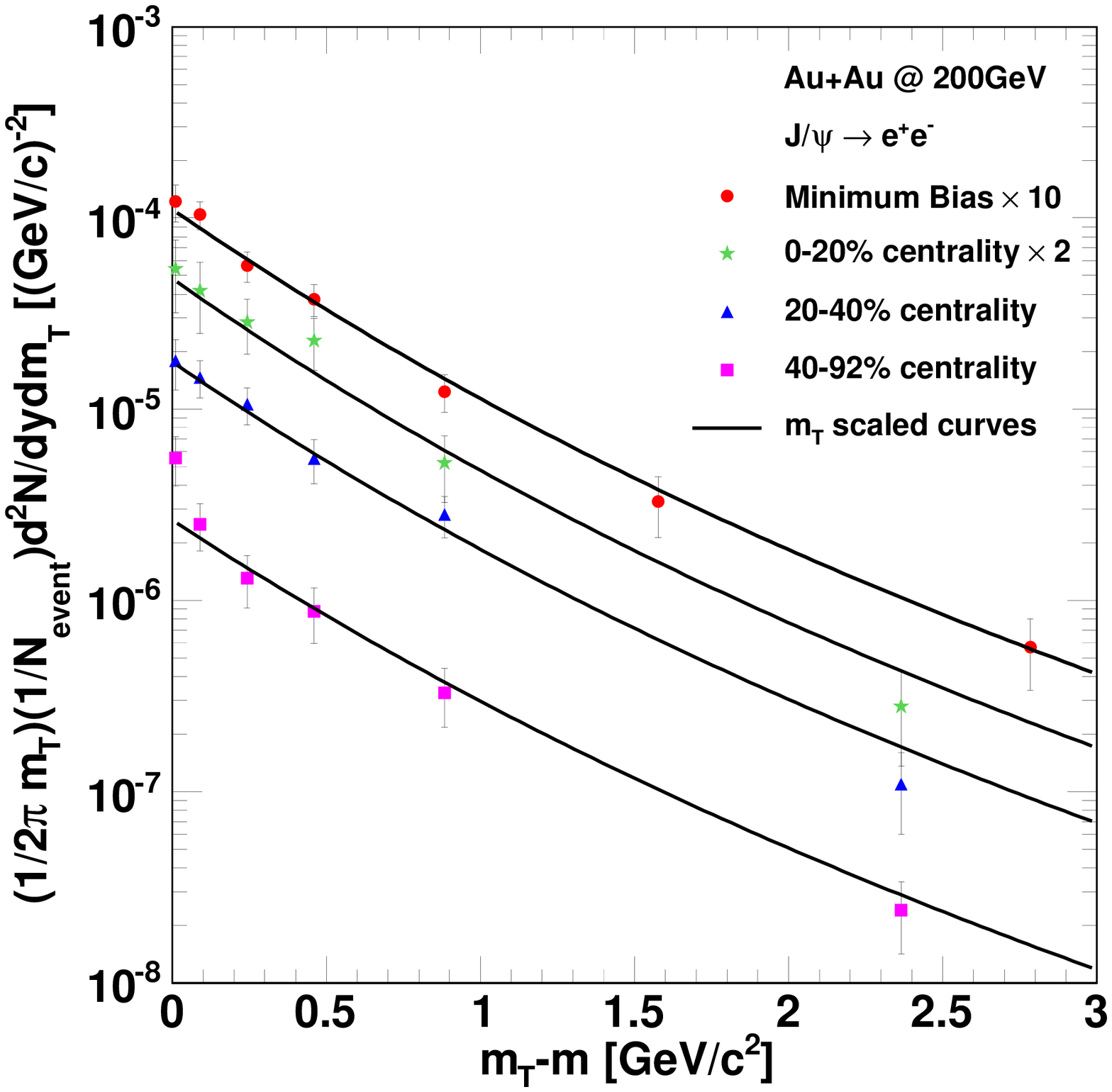}
\caption{(Color online) The invariant yield of measured $J/\psi$ \cite{PPG068} as a function 
of $m_{T}$ in Au+Au system for different centralities. The solid lines are obtained 
using $m_{T}$ scaling; the relative normalization has been used to fit the measured spectra. }
\label{auau14}
\end{figure}

  Figure~\ref{auau10} shows the invariant yield of measured 
neutral \cite{PPG080} and charged pions \cite{PPG026}
as a function of $m_{T}$ in Au+Au system for different centralities. 
The solid lines are the Hagedorn function fits, the parameters of which 
are given in Table \ref{auaucentPar}.
  Figure~\ref{auau11} shows the invariant yield of measured K$^\pm$ \cite{PPG026} as a function 
of $m_{T}$ in Au+Au system for different centralities. The solid lines are obtained 
using $m_{T}$ scaling from the corresponding centrality data of pions.
The relative normalization has been used to fit the measured spectra. 
Figure~\ref{auau12} shows the invariant yield of measured 
$\eta$ (open symbols \cite{PPG051}, solid symbols \cite{PPG115})
as a function of $m_{T}$ in Au+Au system for different centralities. 
Figure~\ref{auau13} shows the invariant yield of measured $\phi$ \cite{PPG096} as a function 
of $m_{T}$ in Au+Au system for different centralities. 
Figure~\ref{auau14} shows the invariant yield of measured $J/\psi$ \cite{PPG068} as a function 
of $m_{T}$ in Au+Au system for different centralities. 

  In case of $\eta$ the data of all centralities are very well reproduced by $m_T$ scaled pion data.
The ratios obtained also remain approximately same.
  In case of kaons, the shapes of the peripheral data are reproduced by $m_T$ scaled pion data
very well but for the most central collisions the disagreement between the two is quite evident.
Also the fitted ratios tend to increase as we move from peripheral to central collisions.
 The same seems true for $\phi$ where peripheral data is better reproduced as compared to 
central data and the ratios also increase as the collision becomes more central.  
In case of J/$\psi$ all three centralities seem same both by shape and relative ratio to the pions.
At least from the present data they can not be distinguished.

\section{conclusion}

 In summary, we parameterize experimentally measured pion spectra 
and then obtain the spectra of other light mesons using a property known as $m_T$ scaling. 
  The $m_{T}$  scaled spectra for each meson is compared with experimental data for
p + p, d + Au and Au+Au systems at $\surd s_{NN} $ = 200 GeV.  Their 
fitted relative normalization gives meson to pion ratio. 
These ratios would be useful to obtain the hadronic decay contribution in 
photonic and leptonic channels.
 The shape of the derived $m_T$ scaled spectra very well reproduce the 
measured spectra and fitted meson to pions are in agreement with the ratios available
in literature for p+p system.
 We show that the ratios of the $\eta$ and $\omega$ $m_T$ spectra with 
pion $m_T$ spectra obtained for p+p collisions can be used for the case of d+Au and Au+Au systems.
 It is observed that the particles with charm contents behave differently from pions for
both the d+Au and Au+Au systems.  
  The particles with strange contents behave similar as pions in d+Au system but
behave differently from pions in Au+Au systems. 
  The more  detailed centrality analysis of Au+Au collision reveal that 
in case of $\eta$ the data of all centralities are very well reproduced by $m_T$ scaled pion data.
The ratios obtained also remain approximately same. For particles, kaon and $\phi$,
peripheral data is better reproduced as compared to central data and the ratios also increase as 
the collision becomes more central.  
  In case of J/$\psi$, all three centralities seems same both by shape and relative ratios to the pions
within the errors.

\section{Acknowledgements}
  We acknowledge the financial support from Board of Research in Nuclear 
Physics (BRNS) for this project.

\begin{thebibliography}{500}

\bibitem{WP} I. Arsene et. al. (BRAHMS), Nucl. Phys. A{\bf 757}, 1 (2005); 
 B. B. Back et. al. (PHOBOS), Nucl. Phys. A{\bf 757}, 28 (2005); 
 J. Adams et. al. (STAR), Nucl. Phys. A{\bf 757}, 10 (2005); 
 K. Adcox et. al. (PHENIX), Nucl. Phys. A{\bf 757}, 184 (2005). 


\bibitem{RAA} K. Adcox et al. (PHENIX), Phys. Rev. Lett. {\bf 88}, 022301 (2001). 

\bibitem{HADPHEN} S. S. Adler et al. (PHENIX), Phys. Rev. Lett. {\bf 94}, 082302 (2005).

\bibitem{PHOPHEN} A. Adare et al. (PHENIX), Phys. Rev. Lett. {\bf 104}, 132301 (2010). 
  
\bibitem{DIELEC} A. Adare et al. (PHENIX), Phys. Rev. C{\bf 81}, 034911 (2010). 

\bibitem{SINGLE} M. L. Mangano et al., Nucl. Phys. B{\bf 405}, 507 (1993).

\bibitem{COCK} A. Adare et al. (PHENIX), Phys. Rev. Lett. {\bf 98}, 172301 (2007). 

\bibitem{WA80} R. Albrecht et. al.(WA80 collaboration), Phys. Lett. B{\bf 361}, 14 (1995);
                        arXiv:hep-ex/9507009.

\bibitem{HAG1} R. Hagedorn,Rev. del Nuovo Cim. {\bf 6N 10}, 1 (1984).
  
\bibitem{HAGEFACT} R. Balnkenbecler and S.J. Brodsky, Phys. Rev. D{\bf 10}, 2973 (1974).
  
\bibitem{UA1} C. Albajar et al. (UA1), Nucl. Phys. B{\bf x }, 1 (1989).

\bibitem{PPG077} A. Adare et al. (PHENIX), arXiv:1005.1627(2010).
 
\bibitem{PPG099} A. Adare et al. (PHENIX), Phys. Rev. D{\bf 83}, 052004 (2011).
    
\bibitem{SEANA} S. Kelly (PHENIX), J. Phys. G{\bf 30}  S1189 (2004).


\bibitem{PPG065} A. Adare et al. (PHENIX), Phys. Rev. Lett. {\bf 97}, 252002 (2006).

\bibitem{PPG055} S. S. Adler et al. (PHENIX), Phys. Rev. C{\bf 75}, 024909 (2007). 

\bibitem{PYTHIA} T. Sjoestrand, S. Mrenna, and P. Skands, ¡°PYTHIA 6.4 physics and manual¡±, JHEP 05, 026 (2006), arXiv:hep-ph/0603175.

\bibitem{RATOMEGA} S. S. Adler et al. (PHENIX), Phys. Rev. C{\bf 75}, 051902 (2007).

\bibitem{VICTOR} V. Ryabov et al., Nucl. Phys. A{\bf 774} 735 (2006).



\bibitem{RATPHI} K. Adcox et al. (PHENIX), Phys. Rev. Lett {\bf 88}, 192303 (2002). 

\bibitem{PPG063} A. Adare et al. (PHENIX), Phys. Rev. D{\bf 76}, 051106(R) (2007).


\bibitem{PPG030} S. S. Adler et al. (PHENIX), Phys. Rev. C{\bf 74}, 024904 (2006). 


\bibitem{PPG115} A. Adare et al. (PHENIX), Phys. Rev. C{\bf 82}, 011902 (2010). 


\bibitem{PPG096} A. Adare et al. (PHENIX), Phys. Rev. C{\bf 83}, 024909 (2011). 

\bibitem{PPG069} A. Adare et al. (PHENIX), Phys. Rev. Lett. {\bf 98}, 232002 (2007).

\bibitem{PPG097} A. Adare et al. (PHENIX), Phys. Rev. D{\bf 82}, 012001 (2010). 

\bibitem{PPG064} S. S. Adler et al. (PHENIX), Phys. Rev. C{\bf 75}, 051902 (2007).


\bibitem{PPG044} S. S. Adler et al. (PHENIX), Phys. Rev. Lett. {\bf 98}, 172302 (2007).

\bibitem{PPG038} S. S. Adler et al. (PHENIX), Phys. Rev. Lett. {\bf 96}, 012304 (2006). 

\bibitem{PPG051} S. S. Adler et al. (PHENIX), Phys. Rev. Lett. {\bf 96}, 202301 (2006). 


\bibitem{PPG080} A. Adare et al. (PHENIX), Phys. Rev. Lett. {\bf 101}, 232301 (2008).


\bibitem{PPG026} S. S. Adler et al. (PHENIX), Phys. Rev. C{\bf 69}, 034909 (2004).

\bibitem{PPG068} A. Adare et al. (PHENIX), Phys. Rev. Lett. {\bf 98}, 232301 (2007).

\bibitem{PPG118} A. Adare et al. (PHENIX), arXiv:1105.3467v1 (2011).





\end{thebibliography}
\end{document}